\documentclass[seceq,preprint]{ptptex}

\usepackage{amsmath,amsfonts,latexsym,amssymb}
\usepackage{verbatim,amsthm,graphicx}
\usepackage{wrapft}

\usepackage{mathrsfs}

\def\linebreak{\hfill\break}

\def\bra<#1|{\langle #1\rvert}
\def\ket|#1>{\lvert#1 \rangle}
\def\braket<#1|#2>{\langle #1|#2 \rangle}

\def\pfrac#1#2{\left(\frac{#1}{#2}\right)}
\def\const{\text{const}}
\def\otop#1{\hbox{$#1\kern-0.1em$\llap{\hbox{\raise1.7ex\hbox{$\scriptstyle\circ$}}}} }

\def\inpare#1{\left(#1\right)}
\def\bigpare(#1){\left(#1\right)}
\def\inrbra#1{\left\{ #1 \right\}}
\def\insbra#1{\left[ #1 \right]}

\def\bigbra[#1]{\left[ #1 \right]}

\def\cases#1{\left\{ \begin{array}{ll}#1\end{array}\right.}

\def\tend{\rightarrow}
\def\then{\Rightarrow\quad}
\def\equivalent{\quad\Leftrightarrow\quad}
\def\therefore{\mbox{\setbox0=\hbox{X}\hbox{$\ldotp$}\raise0.7\ht0\hbox{$\ldotp$}\hbox{$\ldotp$}} \quad }
\def\because{\mbox{\setbox0=\hbox{X}\raise0.7\ht0\hbox{$\ldotp$}\hbox{$\ldotp$}\raise0.7\ht0\hbox{$\ldotp$}}\kern0pt }

\def\r#1{{\rm #1}}

\def\bm#1{\boldsymbol{#1}}

\def\ZR{{{\mathbb Z}}}

\def\RF{{{\mathbb R}}}

\def\maps{\rightarrow}

\def\SetDef#1#2{\def\mstr{\vphantom{#1#2}}
 \left\{\mstr #1 \,\right| \,\left. \mstr #2 \right\}}
\def\upin{\hbox{\setbox0=\hbox{$\cup$} \vrule width 0.05 \wd0 height \ht0
 depth 0pt \kern - 0.5\wd0 \box0 }}
\def\Frac(#1/#2){\left(\frac{#1}{#2}\right)}

\def\Tr{{\rm Tr}}

\def\Tp#1{\,{}^t\! #1}

\def\sdprod{\mathrel{{\setbox0=\hbox{$\displaystyle\times$}\lower0.3\wd0\hbox{$\stackrel{\box0}{\scriptstyle\sim}$}}}}

\def\w{\wedge}

\def\Order#1{\r{O}\!\left(#1\right)}
\def\Tdot#1{{{#1}^{\hbox{.}}}}
\def\Tddot#1{{{#1}^{\hbox{..}}}}

\def\pd{\partial}

\def\dual{{}*\! }

\def\Eq#1{\begin{equation} #1 \end{equation}}

\def\Eqr#1{\begin{eqnarray} #1 \end{eqnarray}}

\def\Eqrsub#1{\begin{subequations}\Eqr{#1}\end{subequations}}
\def\Eqrsubl#1#2{\begin{subequations}
  \expandafter\ifx\csname Rlabel\endcsname \relax \label{#1}
  \else \Rlabel{#1} \fi \Eqr{#2}\end{subequations}}
\def\Bitm{\begin{itemize}}
\def\Eitm{\end{itemize}}
\def\Blist#1#2{\begin{list}{#1}{\parsep=0pt \itemsep=0pt%
  \listparindent=0pt #2}}
\def\Elist{\end{list}}
\long\def\ignore#1#2{\def\ignoreflag{#1}\long\def\tmptext{#2}
  \ifnum\ignoreflag>1 #2 \fi}

\preprintnumber[3cm]{hep-th/0605179\\ YITP-06-22\\ KU-TP 002}
\title{\large Time-Dependent Supersymmetric Solutions in M-Theory \\
and the Compactification-Decompactification Transition}

\author{Hideo {\sc Kodama}$^{1,}$%
\footnote{E-mail: kodama@yukawa.kyoto-u.ac.jp}
and
Nobuyoshi {\sc Ohta}$^{2,}$%
\footnote{E-mail: ohtan@phys.kindai.ac.jp}
}
\inst{$^1$Yukawa Institute for Theoretical Physics, Kyoto University \\
Kyoto  606-8502, Japan \\
$^2$Department of Physics, Kinki University, Higashi-Osaka\\
Osaka 577-8502, Japan
}

\abst{
We show that the diagonal light-like solution
with 16 supersymmetries in eleven-dimensional supergravity derived
in our previous paper 
can be generalised to
non-diagonal solutions preserving the same number  of
supersymmetries. Although the metric coefficients of this generalised
light-like solution depend on only a single null coordinate, this
class of solutions contains a subclass equivalent to the
class of solutions found by Bin Chen that are dependent on the
spatial coordinates.
Utilising these solutions, we construct toroidally compactified
solutions that smoothly connect a static compactified region with a
dynamically decompactifying region along a null hypersurface.
}

\begin{document}
\maketitle

\section{Introduction}

In cosmology based on higher-dimensional unified theories, the dynamics
of moduli associated with compactification are of crucial importance in
constructing a realistic universe model as well as in investigating
the early evolution of the universe.

This problem is closely related to the stability of vacua. In general,
it is often implicitly assumed that a supersymmetric vacuum is
stable. This is generally correct for a supersymmetric theory
without gravity. However, in supergravity, a supersymmetric vacuum
can become unstable with respect to small perturbations if it belongs
to a continuous set of degenerate vacua, that is, if there exist
(massless) moduli degrees of freedom. In general, such instabilities are
associated with dynamically increasing supersymmetry breaking, as in
the case of the size-modulus instability for vacua with flux in
warped compactifications of type IIB supergravity%
\cite{Gibbons.G&Lu&Pope2005,Kodama.H&Uzawa2005,Kodama.H&Uzawa2006}.
Instabilities of this type affect the expansion law of the universe
significantly and often produce naked singularities or a big-rip
singularity\cite{Gibbons.G&Lu&Pope2005,Kodama.H&Uzawa2005}. There
can be other types of instabilities that connect two different
supersymmetric vacua and are associated with topological change of the
internal space\cite{Candelas.P&Green&Hubsch1989,Greene.B1997A},
such as the flop transition\cite{Brandle.M&Lukas2003,Jarv.L&Mohaupt&Saueressig2003}
and the conifold transition\cite{Lukas.A&Palti&Saffin2005,%
Mohaupt.T&Saueressig2005a,Mohaupt.T&Saueressig2005A}.
These types of transitions are also associated with supersymmetry
breaking or singularities.

In the treatment of a singularity appearing in our spacetime,
the usual approach based on general relativity offers little insight.
It is here that superstring theories and M-theory are expected to be useful.
In fact, recent studies of time-dependent supersymmetric solutions
in string theories and M-theory have cast new light on this problem.%
\cite{time1,Craps.B&Sethi&Verlinde2005,bb2,bb4,Chen.B2006,%
Ishino.T&Kodama&Ohta2005,bb8,bb9,bb10,bb11,bb12,bb13,bb14,%
Chen.H&Chen2006A,Ishino.T&Ohta2006A}
In particular, various time-dependent solutions with partial
supersymmetry have been found and used to study the nature of
singularities in the context of cosmology. Solutions with partial
supersymmetry are important because they allow us to investigate
nonperturbative regions of our spacetime. In this case, it has been
argued that such a singularity can be described perturbatively
in terms of dual matrix theories. It would be interesting to explore
this possibility further and obtain more insight
into singularities, instabilities and other aspects of such
time-dependent solutions.

The main purpose of the present paper is to point out that there
exists another type of instability in M-theory, that is, a dynamical
instability of extra dimensions described by an exactly
supersymmetric solution. Because supersymmetric solutions in
supergravity always have time-like or null Killing vectors%
\cite{Tod.K1983,Tod.K1995,Gauntlett.J&&2003,Gauntlett.J&Gutowski2003,%
Gauntlett.J&Pakis2003,Gauntlett.J&Gutowski&Pakis2003,%
Hackett-Jones.E&Smith2004,Gran.U&Gutowski&Papadopoulos2005},
dynamical phenomena can be described by supersymmetric solutions
only when they have null Killing vectors. Recently, we found a
class of such solutions with 16 supersymmetries in
M-theory with a non-trivial 4-form
flux\cite{Ishino.T&Kodama&Ohta2005}. These light-like solutions are
represented by a diagonal metric whose components depend on only a
single null coordinate, $u$. In the present paper, we slightly
generalise these solutions to a class with a non-diagonal metric
preserving the spatial-coordinate independence and study their
toroidal compactification. In particular, we explicitly construct
the compactification of supersymmetric regular light-like solutions for
which the internal space is static in the region of the
spacetime satisfying $u<0$  and dynamically expands as
$u\tend\infty$. We also point out that solutions of this type can be
used to construct a supersymmetric solution that describes the
creation and expansion of a stable compactified region in an
asymptotically decompactifying background.

In this construction of dynamical compactification, it is not
essential that the metric be non-diagonal. However, with this
generalisation to non-diagonal solutions, we can investigate
the problem in a quite wide class of solutions with 16 supersymmetries,
in spite of the fact that the metric is independent of the spatial
coordinates. In fact, we show that through
appropriate coordinate transformations, the apparently more general
solutions with spatial-coordinate dependence found by Bin
Chen\cite{Chen.B2006} can be mapped to a subclass of our generalised
light-like solutions.

The present paper is organised as follows. In the next section, we
show that the diagonal solutions we derived previously can be
generalised to non-diagonal solutions preserving 16 supersymmetries.
Then, we prove that Bin Chen's solutions can be transformed into a subclass of our generalised light-like solutions. Furthermore,
we illustrate this point explicitly for the Kowalski-Glikman solution\cite{Kowalski-Glikman.J1984a} with 32
supersymmetries. In \S3, we study basic geometrical properties
of our solutions. In particular, we determine the common maximal
isometry group of our class of solutions. We also comment on some
exceptional solutions with higher symmetries. With these
preparations, we investigate toroidal compactification of our
solutions and construct dynamically decompactifying solutions in
\S4. Section 5 is devoted to a summary and discussion.

\section{Light-like solutions in M-theory}

\subsection{Generalised light-like solutions}

The bosonic sector of the $D=11$ minimal supergravity consists of a
spacetime metric $g_{MN}$ and a 4-form flux $F_4$. Their field
equations are given by
\Eqrsub{
&& R_{MN} = \frac{1}{12}F_{M P_1P_2 P_3}F_N{}^{P_1 P_2
P_3}-\frac{1}{6}F_4\cdot F_4 g_{MN},
\label{EinsteinEq:general}\\
&& d F_4=0,\quad d\dual F_4 + \frac{1}{4}F_4\w F_4=0.
\label{FormFieldEq:general}
}
Here and in the following, we define the inner product of two $p$-forms as
\Eq{
\alpha_p \cdot \beta_p =\frac{1}{p!}\alpha_{M_1\cdots M_p}\beta^{M_1\cdots M_p}.
}
%

Let us consider the spacetime whose metric is given by
\Eq{
ds^2= -2 du dv + g_{ij}(u) dx^i dx^j.
\label{GLLS:SCI:metric}
}
Here, $g_{ij}(u)$ is a general non-degenerate matrix depending on
only the coordinate $u$. With respect to the null frame,
\Eqrsubl{GLLS:SCI:NullFrame}{
&& \theta^-=-du,\quad \theta^+=dv,\quad \theta^I=\theta^I_i(u)
dx^i,\\
&& e_-=-\pd_u,\quad e_+=\pd_v,\quad e_I=e_I^i(u) \pd_i,
}
where $\theta^I_i e_J^i=\delta^I_J$, the non-vanishing components of
the Riemann curvature tensor are given by
\Eq{
R_{-i-j}=\inpare{-\frac{1}{2}\ddot{g} + \frac{1}{4}\dot{g}g^{-1}\dot{g}}_{ij}.
\label{GLLS:SCI:Riemann}
}
Throughout this paper, a dot represents differentiation with respect
to $u$.

{}From the above, it follows that the components of the Ricci tensor
other than $R_{--}$ vanish. Hence, from the Einstein equations, we
obtain
\Eqrsub{
&& 0\equiv R_{++}=\frac{1}{12}F_{+ijk}F_+{}^{ijk},\\
&& 0\equiv R_{+-}=\frac{1}{6}\inpare{F_{+-ij}F_{+-}{}^{ij}
     +\frac{1}{4!}F_{ijkl}F^{ijkl} }.
}
These equations are equivalent to
\Eq{
F_{+-ij}=F_{+ijk}=F_{ijkl}=0.
}
Under these conditions, $F_4\cdot F_4$ vanishes, and  we can easily see
that the Einstein equations reduce to the single equation
\Eq{
R_{--}\equiv \Tr\inpare{ -\frac{1}{2}g^{-1}\ddot{g}
      + \frac{1}{4}g^{-1}\dot{g}g^{-1}\dot{g}}
      =\frac{1}{12} F_{-ijk}F_-{}^{ijk}.
\label{R--}
}
%

Now, the 4-form flux can be written
\Eq{
F_4= \frac{1}{3!}F_{-ijk} \theta^- \w dx^i \w dx^j \w dx^k
   = - du \w F_-,
\label{GLLS:F4}
}
where
\Eq{
F_- = \frac{1}{3!}F_{-ijk}dx^i \w dx^j \w dx^k.
}
Hence, the field equation $dF_4=0$ yields
\Eq{
\pd_v F_-=0,\quad
d_x F_-=0.
}
Because $F_4\wedge F_4=0$ in the present case, the remaining field
equation for $F_4$ becomes $d\dual F_4=0$. For a $p$-form $\omega_p$
on $X_9 \ni (x^i)$, we have
\Eq{
 \dual (du \w \omega_p) = (-1)^{10-p} du \w \dual_X \omega_p,
}
where $\dual_X$ denotes the Hodge dual on $X_9$ with respect to the metric
$g_{ij}dx^i dx^j$. Hence, the equation $d\dual F_4=0$ is equivalent to
\Eq{
d_x \dual_X F_-=0.
}
Thus, $F_-$ should be a $v$-independent harmonic 3-form on the space $X_9$.
In particular, if  we require that the curvature of the spacetime be
bounded everywhere, \eqref{R--} is consistent only when $F_{-ijk}$
is independent of $v$ and $x^i$. Hereafter, let us
assume this. The general regular solution to
\eqref{EinsteinEq:general} and \eqref{FormFieldEq:general}
with the metric form \eqref{GLLS:SCI:metric} is then given by a set
of functions of $u$, which we denote $g_{ij}(u)$ and $F_{-ijk}(u)$,
satisfying \eqref{R--}. This solution reduces to the
solution found in Ref. \citen{Ishino.T&Kodama&Ohta2005} when
$g_{ij}(u)$ is diagonal.

\subsection{Supersymmetry}
\label{subsec:supersymmetry}

In the Killing spinor equation
\Eq{
(\nabla_M + T_M)\epsilon=0,
}
with
\Eq{
\nabla_M = \pd_M + \frac{1}{4}\omega_{ab M}\Gamma^{ab},
}
and
\Eq{
T_M = \frac{1}{288}\inpare{\Gamma_M{}^{P_1\cdots P_4}
     -8\delta^{P_1}_M \Gamma^{\cdots P_4}}F_{P_1\cdots P_4},
}
%
the connection coefficients for the metric \eqref{GLLS:SCI:metric} with
respect to the null frame \eqref{GLLS:SCI:NullFrame} are given by
\Eqrsub{
&& \omega_{++}=\omega_{+-}=\omega_{--}=\omega_{+I}=0,\\
&& \omega_{-I}= h_{IJ} \theta^J,\quad \omega_{IJ}= \Omega_{IJ} \theta^-,
}
with
\Eq{
h_{IJ} = \frac{1}{2}e_I^i e_J^j \dot g_{ij},\quad
\Omega_{IJ} = e^i_{[J} \dot \theta_{I]i},
}
%
and the matrices $T_M$ are expressed as
\Eqrsub{
 T_- &=& \frac{1}{72}(\Gamma^+\Gamma^- - 3)\Gamma^{IJK} F_{-IJK},\\
 T_+ &=& 0,\\
 T_I &=& \frac{1}{72}\inpare{ -\Gamma_I{}^{JKL}F_{-JKL}
         +6 \Gamma^{JK}F_{-IJK} } \Gamma^-,
}
where $\Gamma^{\pm}=(\Gamma^{10} \pm \Gamma^0)/\sqrt{2}$. 
Therefore, the Killing spinor equation can be written
\Eqrsub{
&& \insbra{ \pd_- + \frac{1}{4}\Omega_{IJ}\Gamma^{IJ}
    +\frac{1}{72}(\Gamma^+\Gamma^- -3)\Gamma^{IJK}F_{-IJK} } \epsilon=0,\\
&& \pd_+\epsilon =0,\\
&&
 \pd_I \epsilon + \insbra{ -\frac{1}{2}h_{IJ}\Gamma^J
    +\frac{1}{72}\inpare{-F_{-JKL}\Gamma_I{}^{JKL}
    +6 F_{-IJK}\Gamma^{JK} } } \Gamma^-\epsilon
  = 0.
}
Here, note that the quantities $\Gamma^{I\cdots J}$ are constant
matrices and that $h_{IJ}, \Omega_{IJ}$ and $F_{-IJK}$ are functions
of only $u$.

Now, let $\zeta$ be a constant spinor satisfying
\Eq{
\Gamma^- \zeta=0
}
and $\epsilon=\epsilon(u)$ be the solution to the ordinary
differential equation
\Eq{
\pd_u \epsilon= \inpare{\frac{1}{4}\Omega_{IJ}\Gamma^{IJ}
     -\frac{1}{24}\Gamma^{IJK}F_{-IJK} } \epsilon
}
with the initial condition $\epsilon(u_0)=\zeta$. Then, from the
relations $\Gamma^-\epsilon(u_0)=0$ and
\Eq{
\pd_u \Gamma^- \epsilon= \inpare{\frac{1}{4}\Omega_{IJ}\Gamma^{IJ}
     +\frac{1}{24}\Gamma^{IJK}F_{-IJK} } \Gamma^-\epsilon,
}
if follows that $\Gamma^-\epsilon =0$ holds for
any value of $u$. This implies that $\epsilon=\epsilon(u)$
is a solution to the above Killing spinor equation for any
constant spinor $\zeta$. Therefore, our generalised
light-like solution always has at least 16 supersymmetries.

\subsection{Relation to Bin Chen's solution}

Recently, Bin Chen found a class of solutions with 16
supersymmetries in M-theory by generalising the plane wave
solution\cite{Chen.B2006}. In an appropriate gauge, his
solution takes the form
\Eqrsub{
&& ds^2=-2du d\bar v + C(y,y) du^2 + E(y,dy) du + D(dy,dy),
\label{BinChen:metric}\\
&& F_4= F_{-123}(u) \theta^- \w dy^1 \w dy^2 \w dy^3,
\label{BinChen:F4}
}
where $C$ and $D$ are quadratic forms, and $E$ is a
bilinear form. The metric coefficients $C_{ij}, D_{ij}$
and $E_{ij}$ are functions of only $u$. In Bin Chen's
paper, it is further assumed that $D$ is diagonal. The quantity
$F_{-123}$ is determined in terms of these metric coefficients by
the Einstein equations, as in our case.

This family of solutions may appear to be more
general than ours or, at least, to contain a class of solutions that
is not included in our class, because the metric
\eqref{BinChen:metric} allows explicit dependence on the
spatial coordinates.  However, this is not the
case. On the contrary, we can show that in fact Bin Chen's solutions
constitute a subclass of our class, at least locally.

To demonstrate the above assertion, let us consider the coordinate
transformation
\Eq{
 u= u,\quad
 v= \bar v + A(y,y)+ L y + m,\quad
 x=B y + N,
\label{GLLS:trf}
}
where $A$ is a quadratic form depending on $u$, $B$ is a
matrix function of $u$, $L$ and $N$ are vector functions
of $u$, and $m$ is a function of $u$. In order for the
metric \eqref{BinChen:metric} to be transformed into the
form \eqref{GLLS:SCI:metric} by this transformation, the
following equations must be satisfied:
\Eqrsubl{BinChenbyGLLS}{
&& C= -2\dot A + \dot B^T g \dot B,\\
&& 0= -\dot L + \dot B^T g \dot N,\\
&& 0= -2\dot m + g(\dot N, \dot N),\\
&& E=-4A + 2\dot B^T g B,\\
&& 0=-L + B^T g \dot N,\\
&& D= B^T g B.
}
{}From these, we obtain 
$L$, $m$ and $A$ in terms of other quantities:
\Eqrsub{
&& L=B^T g \dot N,
\label{LLS:Gen2:L}\\
&& \dot m= \frac{1}{2} g(\dot N, \dot N),
\label{LLS:Gen2:m}\\
&& A= \frac{1}{2} (\dot B^T g B)_s - \frac{1}{4}E_s.
\label{LLs:Gen2:A}
}
Inserting these expressions into \eqref{BinChenbyGLLS}, we obtain
\Eqrsubl{LLS:Gen2}{
&& \Tdot {(g \dot N)}=0,
\label{LLS:Gen2:C1}\\
&& E_a= -2 (B^T g \dot B)_a,
\label{LLS:Gen2:C2}\\
&& 2C - \dot E_s= -2 (B^T \Tdot{(g \dot B)})_s,
\label{LLS:Gen2:C3}\\
&& D = B^T g B,
\label{LLS:Gen2:C4}
}
where the subscripts $a$ and $s$ indicate the anti-symmetric part
and symmetric part, respectively, of the matrix in question.

In \eqref{LLS:Gen2}, the first and the last relations
can be regarded as decoupled equations determining $N$ and
$g$ in terms of the others quantities. Hence, the only non-trivial
conditions following from these are
\Eqrsub{
&& \dot B= B D^{-1} X,\\
&& X_a=-\frac{1}{2}E_a.\\
&& \hspace {-3mm} \dot X_s-X_s D^{-1} X_s - \frac{1}{2}E_a D^{-1} X_s
    +\frac{1}{2} X_s D^{-1} E_a
    =\frac{1}{2}\dot E_s - C -\frac{1}{4} E_a D^{-1} E_a,
}
where $X$ is a matrix function of $u$ that is defined by the first
equation. The second and third relations are identical to
(\ref{LLS:Gen2:C2}) and(\ref{LLS:Gen2:C3}), after eliminating $g$
using (\ref{LLS:Gen2:C4}). These equations constitute a set of
ordinary differential equations in $u$ for $B(u)$ and $X_s(u)$ and
always have a solution, at least locally, for any given matrix
functions $C(u)$, $D(u)$ and $E(u)$. We do not have to assume that
$D(u)$ is diagonal. Further, it is obvious that the transformation
\eqref{GLLS:trf} yields \eqref{GLLS:F4} from \eqref{BinChen:F4}.
Thus, we have shown that our class of solutions is equivalent to a
class obtained from Bin Chen's class by generalising $D$ to a
generic symmetric matrix and $F_4$ to the form \eqref{GLLS:F4}.

Here, note that the metric matrix $g_{ij}(u)$ can always be
diagonalised through a  coordinate transformation of the form
\eqref{GLLS:trf}, and the metric \eqref{GLLS:SCI:metric} can be put
into the form \eqref{BinChen:metric}, if we ignore the 4-form flux
$F_4$. However, it is in general impossible to put $F_4$ of the
generic form \eqref{GLLS:F4} into the special form
\eqref{BinChen:F4} assumed by Bin Chen through such a
transformation simultaneously. This implies that our class contains solutions that
are not contained in Bin Chen's class.

\subsection{The Kowalski-Glikman solution}

Eleven-dimensional supergravity has exactly four types of
solutions with 32 supersymmetries, among which the locally
flat spacetime and the Kowalski-Glikman
spacetime\cite{Kowalski-Glikman.J1984a} can be expressed in
light-like forms.  Let us illustrate the general argument
given in the previous subsection by considering this Kowalski-Glikman
solution, which can be expressed as
\Eqrsubl{KGsol}{
&& ds^2= -2du d\bar v - \mu^2 (4 \bm{y}_3^2 + \bm{z}_6^2) du^2
       + d\bm{y}_3^2 +d\bm{z}_6^2,\\
&& F_4= 6\mu \, du \w dy^1 \w dy^2 \w dy^3,
}
where $(x^i)=(\bm{y}_3,\bm{z}_6)$ and $\mu$ is a constant
parameter.

For $E=0$ and $D=I_9$, in the general argument given in the
previous subsection, $X$ becomes a symmetric matrix and
satisfies the equation
\Eq{
\dot X - X^2= -C;\quad
C=-\mu^2 \begin{pmatrix} 4 I_3 & 0 \\ 0 & I_6
\end{pmatrix}.
}
This equation has the diagonal solution
\Eq{
X_{ij}=  a_i \tan(a_i u) \delta_{ij};\quad
a_i=2\mu\ (1\le i\le3),\  \mu\ (4\le i \le 9).
}
The corresponding expressions for $A,B$ and $g$ are
\Eqrsub{
A_{ij} &=& \frac{1}{2}a_i \tan(a_i u) \delta_{ij} ,\\
B^i{}_j &=& \frac{1}{\cos(a_i u)}\delta_{ij} ,\\
g_{ij} &=& \cos^2(a_i u) \delta_{ij} .
}

Thus, we have found that under the coordinate transformation
\Eqrsub{
v &=& \bar v + \mu \tan(2\mu u) \bm{y}_3^2
         +\frac{\mu}{2}\tan(\mu u) \bm{z}_6^2,\\
\bm{x} &=& \inpare{ \frac{1}{\cos(2\mu u)}\bm{y}_3,
               \frac{1}{\cos(\mu u)}\bm{z}_6 },
}
the Kowalski-Glikman solution can be put into the
spatial-coordinate-independent form
\Eqrsubl{KGsol:SCI}{
&& ds^2= -2du dv + \sum_i \cos^2(a_i u) (dx^i)^2,\\
&& F_4= \mu \cos^3 (2\mu u)\, du\w dx^1 \w dx^2 \w dx^3.
}
It is also easy to show that this solution allows 32 Killing spinors.
In fact, they are explicitly given by
\Eq{
\epsilon=\inpare{ 1 + \frac{1}{2} \sum_i x^i a_i e^{a_i u \Gamma^{123}}
\Gamma^{123} \Gamma^i \Gamma^{-} }
\exp\inrbra{-\frac{\mu}{2}(1+\Gamma^-\Gamma^+)
  \Gamma^{123} u} \zeta,
}
where $\zeta$ is an arbitrary constant spinor.

Here, note that the metric in the expression
\eqref{KGsol:SCI} is singular at the values of $u$ where
one of $\cos(a_i u)$ vanishes. As is clear from the
argument, these are not real singularities because
\eqref{KGsol} is not singular there. In fact, from
\eqref{GLLS:SCI:Riemann}, the non-vanishing components of
the Riemann curvature are given by $R_{-i-j}=a_i^2 g_{ij}$
and are finite there. Hence, each regular region of the
expression \eqref{KGsol:SCI} covers only a part of the
maximal extension given by the original expression
\eqref{KGsol}.

\section{Geometry of the solution}

In this section, we study basic geometrical features of
the generalised light-like solution obtained in the
previous section. In particular, we determine the common
isometry group of the solution, i.e., the set of
transformations that are isometries of the solution
for any choice of $g_{ij}(u)$.

\subsection{Isometry group}

The Killing equation for the spacetime
\eqref{GLLS:SCI:metric} is given by
\Eqrsub{
&& \nabla_u \xi_u= \pd_u \xi_u =0,
\label{GLLS:Isom2:KillingEq:uu}\\
&& \nabla_v \xi_v = \pd_v \xi_v=0,
\label{GLLS:Isom2:KillingEq:vv}\\
&& \nabla_u \xi_v + \nabla_v \xi_u=\pd_u \xi_v + \pd_v \xi_u=0,
\label{GLLS:Isom2:KillingEq:uv}\\
&& \nabla_u \xi_i + \nabla_i\xi_u
  =\pd_u\xi_i-\dot{g}_{ik} \xi^k + \pd_i \xi_u=0,
\label{GLLS:Isom2:KillingEq:ui}\\
&& \nabla_v \xi_i + \nabla_i\xi_v
  =\pd_v\xi_i + \pd_i \xi_v=0,
\label{GLLS:Isom2:KillingEq:vi}\\
&& \nabla_i\xi_j+\nabla_j\xi_i=\pd_i\xi_j+\pd_j\xi_i-\dot{g}_{ij}\xi_v=0.
\label{GLLS:Isom2:KillingEq:ij}
}
%
The first three of these equations imply
\Eqrsub{
&& \xi^v= -\xi_u = \kappa (x) v + \eta^v(x),\\
&& \xi^u= -\xi_v = -\kappa (x) u + \eta^u(x).
}
{}From these and \eqref{GLLS:Isom2:KillingEq:vi}, we have
\Eq{
\xi_i = g_{ij}\zeta^j (u,x) - v \pd_i (u \kappa  - \eta^u).
}
Substituting this into \eqref{GLLS:Isom2:KillingEq:ui}, we
obtain
\Eq{
\pd_u\inrbra{\zeta^i-v g^{ij}\pd_j(u\kappa  - \eta^u)}
  - g^{ij} \pd_j(v\kappa  +\eta^v)=0,
}
which is equivalent to the following two equations:
\Eqrsub{
&& \inrbra{\Tdot{(ug^{ij})}+g^{ij}} \pd_j \kappa
  - \Tdot{(g^{ij})} \pd_j \eta^u=0,\\
&& \zeta^i= \eta^i(x) + G^{ij}\pd_j \eta^v.
}
Here, $G^{ij}$ is defined by
\Eq{
G^{ij}:=\int^u_{u_0} du g^{ij}(u),
\label{G:def}
}
with $u_0$ being a constant. Therefore, the last Killing equation,
\eqref{GLLS:Isom2:KillingEq:ij}, can be written
\Eqr{
&& (g_{ik}\pd_j+g_{jk}\pd_i)\eta^k
  + (G_i^k\pd_j + G_j^k\pd_i)\pd_k \eta^v
  \notag\\
&&\quad
  + 2 v\pd_i\pd_j (-u \kappa  + \eta^u)
  +\dot{g}_{ij}(- u \kappa  + \eta^u)=0.
}
In particular, by comparing the term proportional to $v$, we obtain
\Eq{
\pd_i \pd_j \kappa =0,\quad
\pd_i \pd_j \eta^u=0.
}

To summarise, the general solution to the Killing equation,
\eqref{GLLS:Isom2:KillingEq:uu}--\eqref{GLLS:Isom2:KillingEq:ij},
can be expressed as
\Eqrsubl{GLLS:SCI:Killing}{
\xi^v &=& v \kappa  + \eta^v.\\
\xi^u &=& -u \kappa  + \eta^u,\\
\xi^i &=& \eta^i + G^{ij}\pd_j \eta^v
     +v g^{ij}( -u \kappa _j + \eta^u_{j}),
}
where $\kappa _j$ and $\eta^u_{j}$ are constant vectors satisfying
\Eq{
 \Tdot{(u^2 g^{ij})} \kappa _j - u \Tdot{(g^{ij})} \eta^u_{j}=0,
\label{GLLS:KillingEq:reduced1}
}
$\kappa $ and $\eta^u$ are $u$-independent linear functions of $x^i$
given by
\Eq{
\kappa =\kappa _i x^i + \kappa _0,\quad
\eta^u= \eta^u_{i}x^i + \eta^u_{0}
}
and $\eta^k(x)$ and $\eta^v(x)$ are $u$-independent functions of
$x^i$ satisfying
\Eq{
(g_{ik}\pd_j+g_{jk}\pd_i)\eta^k
  +(G_i^k\pd_j + G_j^k\pd_i)\pd_k \eta^v
  +\dot{g}_{ij} ( -u \kappa  + \eta^u)=0.
\label{GLLS:KillingEq:reduced2}
}
%

We can easily show that the reduced Killing equations
\eqref{GLLS:KillingEq:reduced2} and \eqref{GLLS:KillingEq:reduced1}
are satisfied for any choice of $g_{ij}(u)$ when $\kappa
=0,\eta^u=0, \eta^i=\const$ and $\eta^v$ is linear in $x^i$. In this
case, the Killing vectors are parameterised in terms of two constant
vectors, $A^i$ and  $B_i$, and one constant scalar, $C$, as
\Eqrsubl{GenericMaximumKilling}{
\xi^u &=& 0,\\
\xi^v &=& C + B_i x^i,\\
\xi^i &=& A^i + G^{ij} B_j.
}
Actually, this is the maximum common set of Killing vectors for all
possible choices of  $g_{ij}(u)$, as shown in the next subsection.
The corresponding isometry group $G_m$ consists of transformations of the form
\Eqrsub{
u' &=& u,\\
v' &=& v + \frac{1}{2}G^{ij}(u) b_i b_j + b_i x^i + c,\\
x'{}^i &=& x^i + a^i + G^{ij}(u)b_j,
}
where $a=(a^i),b=(b_i)$ and $c$ are constant vectors and a constant
corresponding to $A^i, B_i$ and $C$ in the above expression for the
Killing vector, respectively. If we denote this transformation by
$g=(a,b,c)$, the product of two transformations is expressed as
\Eq{
g' g = (a'+a,b'+b,c'+c + b'\cdot a).
}
In particular, the commutator of two transformations is given by
\Eq{
[g',g]\equiv g' g (g')^{-1} g^{-1}= (0,0, b'\cdot a- b \cdot a').
\label{GLLS:Isom:Commutant}
}
Hence, $G_m$ is a nilpotent group, which becomes identical to the
Heisenberg group in the case in which $a$ and $b$ have single
components\cite{Kodama.H1998,Kodama.H2002}.

To be precise, the maximum common isometry group may contain extra
discrete transformations. Such discrete elements can be
systematically determined by calculating the normaliser group of the
connected component $G_m$, as was done in Refs. \citen{Kodama.H1998}
and \citen{Kodama.H2002} for Bianchi models.
However, we do not investigate this problem in the present paper.

\subsection{Exceptional geometry}
\label{subsec:ExceptionalGeometry}

In this subsection, we show that the isometry group $G_m$ found in
the previous subsection is really the maximum isometry group for
some solutions.
We also give some exceptional solutions which have larger isometry groups.

Here we restrict our consideration to the case of a diagonal metric,
for which we have
\Eq{
g_{ij}=e^{2f_i(u)} \delta_{ij},
}
and the matrix $G$ is diagonal:
\Eq{
G^{ij}=F^i(u) \delta^{ij};\quad
F^i=\int^u_{u_0} du \, e^{-2f_i}.
}
%
In this case, the condition \eqref{GLLS:KillingEq:reduced1} can be written
\Eq{
  (1-u \dot f_i) \kappa_i= -\dot f_i \eta^u_{i}.
}
If $(1-u \dot f_i)\eta_i^u \neq 0$, we can write
$\kappa_i/\eta_i^u = -\dot f_i /(1-u \dot f_i)$. Since the left-hand
side is independent of $u$, the $u$-derivative of the right-hand
side should vanish:
\Eq{
\Tdot{\pfrac{ \dot f_i}{1-u\dot f_i}}
  = \frac{\ddot f_i+(\dot f_i)^2}{(1-u\dot f_i)^2}=0.
}
This implies that $e^{f_i}$ is at most linear in $u$. Including the
remaining case of vanishing $\eta_i^u$, we find that
there exist the following three cases, up to rescaling of $x^i$ by
constants:
\Eqrsub{
e^{f_i}=1 &\then & \kappa_i=0\ \text{with\ arbitrary}\ \eta^u_{i},\label{GLLS:DSCI:metrictype1}\\
e^{f_i}=|u-u_i| &\then& \eta^u_{i}= u_i \kappa_i,
\label{GLLS:DSCI:metrictype2}\\
K_i \not=0 &\then& \kappa_i=\eta^u_{i}=0,
\label{GLLS:DSCI:metrictype3}
}
where
\Eq{
K_i := e^{-f_i} \Tddot {(e^{f_i})}.
}
%

Here, note that $K_i=0$ for the first two cases. Because
\Eq{
R_{-i-j}=- K_i g_{ij}
}
for the diagonal spatial metric, from \eqref{GLLS:SCI:Riemann}, it follows that the spacetime is flat if $K_i=0$ for all $i$. In fact, the  metric
\Eq{
ds^2= -2du dv + \sum_{j}(u-u_j)^2 (dx^j)^2
}
is transformed into
\Eq{
ds^2=-2 du d\tilde v + \sum_j (d\tilde x^j)^2
}
by the transformation
\Eqrsub{
&& \tilde x^j= (u-u_j) x^j,\\
&& \tilde v=v + \frac{1}{2}\sum_j (u-u_j) (x^j)^2.
}
%

Therefore, we can assume without loss of generality that either
\eqref{GLLS:DSCI:metrictype1} or \eqref{GLLS:DSCI:metrictype3} holds for
each $i$. Let $S_1$ be the set of indices corresponding to the first case
and $S_2$ be that corresponding to the second case. Then, as shown in
Appendix \ref{Appendix:ExceptionalKilling}, the general solution to the
Killing equation is given by
\Eqrsubl{GLLS:DSCI:Killing}{
& \xi^u & = - \kappa u + d_0 + \sum_{i\in S_1} d_i x^i,\\
& \xi^v & = \kappa v + C + \sum_{i} B_i x^i
   + \sum_{k\in S_2} \ell_k (x^k)^2
   -\sum_{k,l\in S_2} s_{kl}\omega_{kl} x^k x^l,\\
& \xi^i & = \sum_{j\in S_1} \omega_{ij}x^j + A^i + B_i u + d_i v,\
i\in S_1\\
& \xi^k & = \sum_{l\in S_2} e^{-f_k+f_l} \omega_{kl}x^l
        +(\kappa u -d_0) \dot f_k x^k + A^k + B_k F^k,\
        k\in S_2.
}
Here, $\kappa, d_0, d_i, A^i, B_i$ and $\omega_{ij}=-\omega_{ji}$ are
constants, with  $d_i=0$ if $S_2$ is not empty  and $\omega_{kl}\not=0$
for $k,l\in S_2$ if and only if
\Eq{
2s_{kl}:= e^{f_k+f_l}(\dot f_k- \dot f_l)=\const
\equivalent K_k=K_l.
\label{DLLS:ExceptionalSol:Cond1}
}
Further, $\ell_k$ is a constant related to $f_k$ by \eqref{ellk:def}.
As shown below, $\ell_k\not=0$ if and only if
\Eq{
(\kappa,d_0)\not=(0,0)
\equivalent 2K_k \ddot K_k -3 (\dot K_k)^2=0\ \forall k\in S_2.
\label{DLLS:ExceptionalSol:Cond2}
}
Therefore, it is clear that the Killing vectors are exhausted by
\eqref{GenericMaximumKilling} if $K_i\not= K_j$ ($i\not=j$) and
$2K_k \ddot K_k - 3(\dot K_k)^2\not=0$ for all possible $i,j$ and $k$.

Now, we give some explicit examples of exceptional geometries in the diagonal case. Suppose that for some constants $a,b$ and $\ell$, $f(u)$ satisfies
\Eq{
\Tdot{\inrbra{(au+b)\dot f}}e^{2f}=2l.
\label{DLLS:Exceptional:f}
}
Then, by differentiating this with respect to $u$, we obtain
\Eq{
(au+b)\dot K + 2a K =0
}
where
\Eq{
K:= e^{-f}\Tddot{(e^f)}.
}
When $K\not\equiv 0$ and $a\not=0$, this equation leads to
\Eq{
\Tdot{\pfrac{K}{\dot K}}=-\frac{1}{2}
\quad\equivalent\quad
2K \ddot K - 3(\dot K)^2=0.
\label{DLLS:ExceptionalSol:Cond'}
}
Next for the case in which $a=0$ and $b\not=0$, $K$ must be constant. Hence, \eqref{DLLS:ExceptionalSol:Cond'} is satisfied in this case as well.

Conversely, suppose that \eqref{DLLS:ExceptionalSol:Cond'} is
satisfied. Then, if $\dot K\not\equiv0$, we can easily show that
$f(u)$ satisfies  \eqref{DLLS:Exceptional:f} by tracing the above
argument in the reverse direction. Further, if $K=\const$,
\eqref{DLLS:Exceptional:f} holds if $a=0$ and $b\not=0$. Therefore,
\eqref{DLLS:Exceptional:f} and \eqref{DLLS:ExceptionalSol:Cond'} are
equivalent. This proves the equivalence of $\ell_k \not=0$ and
\eqref{DLLS:ExceptionalSol:Cond2}.

With the help of the above argument, we can easily find metrics
satisfying the condition \eqref{DLLS:ExceptionalSol:Cond2}. For
notational simplicity, we suppress the index $k$ for the time being.

\medskip
\noindent
{\bf i) The case of $\kappa=0, d_0\not=0$:} In this case, $K$  must
be constant. Hence, we have
\Eq{
\Tddot{(e^f)}= \nu^2 e^f,
}
where $\nu^2$ is a real constant. The general solution to this equation is
\Eq{
e^f= D_1 \cosh \nu u + \frac{D_2}{\nu} \sinh \nu u.
}
Here, it is understood that $e^f$ for $\nu=0$ is obtained by taking
the limit $\nu\tend0$ with $D_1$ and $D_2$ constant. The
corresponding value of $\ell$ is given by
\Eq{
2\ell=d_0(D_2^2-\nu^2 D_1^2).
}
Note that the spacetime in  this case is regular, even at the values
of $u$ where $e^{f_k}$ vanishes, and can be extended across the
corresponding null surfaces, because $R_{-i-}{}^j=-\nu_i^2
\delta^j_i$ is regular and finite there.
In fact, the Kowalski-Glikman solution is included in this case.

\medskip
\noindent
{\bf ii) The case of $\kappa \not=0$:} In this case, from the
relation
\Eq{
\frac{\dot K}{K}=-\frac{2\kappa}{\kappa u-d_0},
}
we obtain
\Eq{
\Tddot{(e^f)}= \frac{\nu^2-\frac{1}{4}}{(u-d_0/\kappa)^2} e^f.
}
The general solution to this equation can be written
\Eq{
e^f= (u-d_0/\kappa)^{1/2} \insbra{ D_1\cosh \nu \tilde u
          + \frac{D_2}{\nu} \sinh \nu \tilde u},
}
where
\Eq{
\tilde u=\ln |u-d_0/\kappa|.
}
The corresponding value of $\ell$ is
\Eq{
\ell=\frac{\kappa}{2}(\nu^2 D_1^2 -D_2^2).
}
%

\section{Toroidal compactification}
\label{sec:Compactification}

In this section, we consider toroidal compactification of generalised
light-like solutions of the form
\Eqr{
ds^2 &=& -2du dv + d\bm{y}\cdot d\bm{y} + g(d\bm{z},d\bm{z})
 \notag\\
   &=& -(dx^0)^2 + d\bm{x}\cdot d\bm{x} + g(d\bm{z},d\bm{z}),
\label{GLLS:TC:metric}
}
where $\bm{y}=(x^1,\cdots,x^m)$, $\bm{z}=(x^{m+1},\cdots,x^9)$
and $\bm{x}=(x^{10},\bm{y})$, with $u=(x^0-x^{10})/\sqrt{2}$ and
$v=(x^0+x^{10})/\sqrt{2}$. The quantity
$g(d\bm{z},d\bm{z})=g_{pq}dz^p dz^q$ is an inner product depending
on $u$. We regard the $z$-directions as the extra dimensions and
compactify them. Because $g_{pq}$ depends on $u$, the size of the
compactified extra dimensions changes with $u$ or with $x^0$ and
$x^{10}$. We show that we can construct a solution describing the
compactification-decompactification transition utilising this
feature. We do not specify the 4-form flux at the beginning, apart
from the general constraint \eqref{GLLS:F4} with
$F_{-ijk}=F_{-ijk}(u)$ and the Einstein equation \eqref{R--}
connecting $F_{-ijk}$ and the spacetime metric.

\subsection{Abelian discrete isometry group}

In order to construct a toroidal compactification of our
solution, we have to find an abelian discrete subgroup
$\Gamma$ of the isometry group $G_m$. Let
\Eq{
g_{p}=(\bm{a}, \tilde{\bm{a}}, \bm{b},\tilde{\bm{b}},c)_p, \
p=1,\cdots,9-m
}
be  a set of elements in $G_m$ that represent the transformations
\Eqrsubl{DiscreteIsometry}{
u' &=& u,\\
v' &=& v + \frac{u}{2}\bm{b}^2 + \frac{1}{2}G(\tilde{\bm{b}},\tilde{\bm{b}})
 + \bm{b}\cdot\bm{y} + \tilde{\bm{b}}\cdot\bm{z}+ c,\\
\bm{y}' &=& \bm{y}+ u \bm{b} + \bm{a},\\
\bm{z}' &=& \bm{z} + G \tilde{\bm{b}} + \tilde{\bm{a}}.
}
Here, for simplicity, we have set $u_0=0$ in the
definition of $G^{ij}$ in \eqref{G:def}, and $G=(G^{pq})$
is regarded as either a quadratic form or a linear transformation in
$\tilde{\bm{b}}$ space, depending on the situation.

Then, these transformations commute if and only if we have
\Eq{
\bm{b}_p \cdot \bm{a}_q + \tilde{\bm{b}}_p
\cdot\tilde{\bm{a}}_q=\bm{b}_q \cdot \bm{a}_p + \tilde{\bm{b}}_q
\cdot\tilde{\bm{a}}_p
\quad \forall\,p,q=1,\cdots,9-m,
}
from the commutation relation \eqref{GLLS:Isom:Commutant}.
In this paper, we only consider the case in which
$\bm{b}_p=0$ for simplicity, and we do not require  $\bm{a}_p$ to
vanish for the reason explained below. Then, the above condition on
the commutativity simplifies to
\Eq{
 \tilde{\bm{b}}_p \cdot\tilde{\bm{a}}_q
 =\tilde{\bm{b}}_q \cdot\tilde{\bm{a}}_p
\quad \forall\,p,q=1,\cdots,9-m.
\label{CommutativityCond}
}
When $\tilde{\bm{b}}_p\ (p=1,\cdots,9-m)$ are linearly independent,
this condition can be expressed in terms of the matrices $T=(T^p_q)$
and $B=(B_{pq})$ defined by
\Eqr{
&& \tilde {\bm{a}}_p = T_p{}^q\tilde{\bm{b}}_q,\\
&& B_{pq}=\tilde{\bm{b}}_p \cdot \tilde{\bm{b}}_q,
}
as
\Eq{
TB=B\Tp T.
}
This is equivalent to the condition that $T$ can be expressed in
terms of a symmetric matrix $S$ as
\Eq{
T= S B^{-1}.
}
For example, when both $T$ and $B$ are diagonal, this condition is
satisfied.

When $\bm{b}_p=0$ and the commutativity condition is satisfied, a
generic element of the discrete transformation group
\Eq{
\Gamma:=\SetDef{\prod_p g_p^{n_p}}{n_p\in\ZR}
}
can be expressed as
\Eqr{
&& \prod_{p} g_p^{n_p}=(\bm{a},\tilde{\bm{a}},0,\tilde{\bm{b}},c);\\
&& \bm{a}= \sum_p n_p \bm{a}_p,\quad
   \tilde{\bm{a}}= \sum_p n_p \tilde{\bm{a}}_p,\quad
   \tilde{\bm{b}}= \sum_p n_p \tilde{\bm{b}}_p,
\label{generator:ab}\\
&& c-\frac{1}{2}\tilde{\bm{a}}\cdot \tilde{\bm{b}}
   =\sum_p n_p \inpare{ c_p
     -\frac{1}{2}\tilde{\bm{a}}_p\cdot \tilde{\bm{b}}_p }.
\label{generator:c}
}
In particular, when the vectors $(\bm{a},\bm{b},\tilde{\bm{b}})_p\
(p=1,\cdots,9-m)$ are linearly independent, $\Gamma$ is isomorphic
to a $(9-m)$-dimensional lattice:
\Eq{
\Gamma \cong \ZR^{9-m}.
}
%

\subsection{General behaviour of the solution}

Before studying the compactification of the spacetime
\eqref{GLLS:TC:metric} by the lattice group $\Gamma$ obtained in the
previous subsection, we point out one important feature of the
generalised light-like solution. Let us write the symmetric metric matrix
$g(u)$ in terms of a diagonal matrix $D(u)$ and an orthogonal matrix
$O(u)$ as
\Eq{
g = O^T D O,
}
and define the anti-symmetric matrix $\Omega$ by
\Eq{
\Omega:= \dot O O^T.
}
Then, the Ricci curvature $R_{--}$ can be expressed as
\Eq{
R_{--}= -\sum_p K_p
 - \sum_{p>q}\frac{(D_p-D_q)^2}{2D_p D_q}(\Omega_{pq})^2,
}
where $D_p$ ($p=1,\cdots,9-m$) are the diagonal entries of $D$ and
\Eq{
K_p := D_p ^{-1/2}\Tddot{(D_p^{1/2})}.
}
Hence, from the Einstein equation \eqref{R--}, we obtain
\Eq{
K:=\sum_p K_p \le 0,
}
where the equality holds only when the 4-form flux vanishes and $g$
is diagonal. This implies that $K<0$ in some region if the
4-form flux does not vanish identically. Then, for
\Eq{
\Delta:= (\det g)^{\frac{1}{2n}}= \prod_p D_p^{\frac{1}{2n}},\ n=9-m,
}
we obtain
\Eq{
n \Delta^{-1} \ddot \Delta \equiv K
 -\frac{1}{4}\sum_p (D_p^{-1}\dot D_p)^2
 + \frac{1}{4n}\inpare{\sum_p D_p^{-1}\dot D_p }^2
 \le K \le 0,
}
where the last two equalities hold simultaneously only when $g$ is
diagonal and $\Tddot{(D_p^{1/2})}=0$, i.e., when the spacetime is
locally flat. Hence, if the spacetime is asymptotically flat in the
limit $u\tend \infty$ or $u\tend-\infty$ and the 4-form flux does
not vanish, $\det g$ (and hence some $D_p$) should vanish at a
finite value of $u$.

\subsection{Compactification}

For this general feature, we cannot construct a non-trivial regular
spacetime through the simplest compactification such that $\Gamma$
consists of simple translations in the $z$-direction, i.e.,
$\bm{a}=\bm{b}=\tilde{\bm{b}}=0$, if we require that the size of the
extra dimensions approach a constant in the limit  $u\tend\infty$ or
$u\tend-\infty$.
However, if we consider transformations with $\bm{a}\not=0$ and
$\tilde{\bm{b}}\not=0$, we can construct a regular spacetime that is
compactified on one side of a null plane with constant $u$. For
example, let us consider the solution in which $g$ is represented by
the diagonal matrix $g_{pq}=e^{2f_p} \delta_{pq}$ satisfying the
conditions
\Eq{
e^{f_p} = \cases{ \alpha_p u + \Order{u^3}&\text{for}\ u\sim0, \\
                  1 + \Order{u^{-2}} &\text{for}\ u\tend\infty. }
\label{DLLS:AC}
}
Then, $G^{pq}=F_p \delta_{pq}$ behaves as
\Eq{
F_p \approx \cases{ -\frac{1}{\alpha_p^2 u} + e_p' &\text{for}\ u \sim0,\\
                    u + e_p &\text{for}\ u \tend\infty.}
}
Hence, for $u\tend\infty$, the transformation
$g=(\bm{a},\tilde{\bm{a}},0,\tilde{\bm{b}})\in \Gamma$ can be written
\Eqrsub{
u' &=& u,\\
v' &\approx& v + \frac{1}{2} \tilde{\bm{b}}\cdot(u+e)\tilde{\bm{b}}
             + \tilde{\bm{b}}\cdot \bm{z} + c,\\
\bm{y}' &=& \bm{y} + \bm{a},\\
\bm{z}' &\approx & \bm{z} + (u+e)\tilde{\bm{b}}+ \tilde{\bm{a}},
}
where $e$ is the diagonal matrix $e_p \delta_{pq}$. Clearly, the
size of the compactified $z$-space diverges as $u\tend\infty$ if
$\tilde{\bm{b}}\not=0$.

Next, in the region with $u\sim0$, by the coordinate transformation
\Eqrsub{
&& \bar v = v + \frac{u}{2}(\alpha \bm{z})^2,\\
&& \bar {\bm{z}}= u \alpha \bm{z},
}
where $\alpha$ is the diagonal matrix $\alpha_p \delta_{pq}$, the
spacetime metric can be written
\Eq{
ds^2\approx -2 du d\bar v + d\bm{y}^2 + d\bar{\bm{z}}^2.
}
Thus, the spacetime approaches a Minkowski spacetime as $u\tend0$,
as expected from the discussion in \S
\ref{subsec:ExceptionalGeometry}. We can make it exactly a Minkowski
spacetime near $u=0$ if necessary.

In this region, $g\in \Gamma$ can be written
\Eqrsub{
u' &=& u,\\
\bar v' &=& \bar v + \inpare{ \tilde{\bm{b}}\cdot \alpha e'
    +\tilde{\bm{a}}\cdot \alpha } \bar{\bm{z}}
       + c-\tilde{\bm{a}}\cdot\tilde{\bm{b}}
       -\frac{1}{2}\tilde{\bm{b}}\cdot e'\tilde{\bm{b}}
   +\Order{u},\\
\bm{y}' &=& \bm{y} + \bm{a},\\
\bar{\bm{z}}' &=& \bar{\bm{z}} - \alpha^{-1}\tilde{\bm{b}}
   +\Order{u},
}
where $e'$ is the diagonal matrix $e'_p \delta_{pq}$. Here, if we
choose the generators of $\Gamma$ so that they satisfy
\Eqrsub{
&& \tilde {\bm{a}}_p=-e' \tilde{\bm{b}}_p,\\
&& c_p=\frac{1}{2} \tilde{\bm{b}}_p\cdot \tilde{\bm{a}}_p,
}
the above transformation of $\bar v$ simplifies to
\Eq{
\bar v' = \bar v + \Order{u},
}
from \eqref{generator:ab} and \eqref{generator:c}. These conditions
on the generators are consistent with the commutativity condition
\eqref{CommutativityCond}.

\begin{figure}
\begin{center}
\includegraphics*[width=8cm]{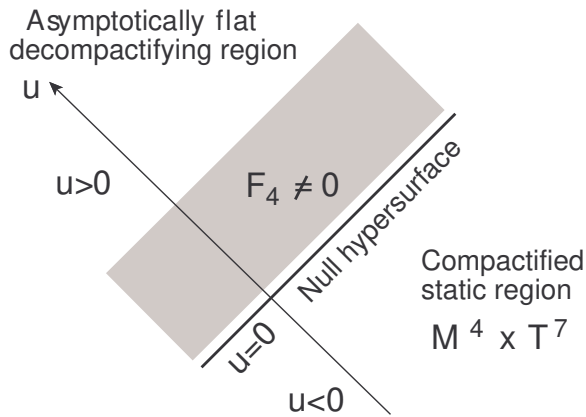}
\caption{Extension of a decompactifying solution with $m=2$ to a static
compactified region across the null hypersurface $u=0$.}
\label{fig:GLLS}
\end{center}
\end{figure}

{}From the above results, we can conclude that the compactification in
this region is identical to the standard toroidal compactification,
apart from a tilt of the extra-dimension direction toward the 
$\bm y$-direction in the
case $\bm{a}\not=0$. This tilt can be eliminated through a linear
coordinate transformation. Further, in terms of the coordinate
system $(u,\bar v, \bm{y},\bar{\bm{z}})$,  we can extend this
compactification to the region $u<0$, as illustrated in
Fig.~\ref{fig:GLLS}. For example, if we assume that the 4-flux
vanishes and the spacetime is exactly locally flat there, the action
of $g\in\Gamma$ is identical to the simple spatial
translation
\Eq{
u'=u,\quad
\bar v'=\bar v,\quad
\bm{y}'=\bm{y}+\bm{a},\quad
\bar{\bm{z}}'=\bar{\bm{z}}-\alpha^{-1}\tilde{\bm{b}}.
}
Hence, the compactified spacetime is exactly the product of the
$(m+2)$-dimensional Minkowski spacetime and a flat torus $T^{9-m}$
with constant sizes.

Finally, let us examine the regularity of this compactified
spacetime in the region $u>0$. For the above parameter choice, $g\in
\Gamma$ can be written
\Eqrsub{
u' &=& u,\\
v' &=& v + \frac{1}{2}\tilde{\bm{b}}\cdot \hat G
   \tilde{\bm{b}}+\tilde{\bm{b}}\cdot\bm{z},\\
\bm{y}' &=& \bm{y} + \bm{a},\\
\bm{z}' &=& \bm{z} + \hat G \tilde{\bm{b}},
}
where
\Eq{
\hat G^{pq} = (F_p - e'_p)\delta_{pq}.
}
Note that this transformation preserves the quantity
\Eq{
v' -\frac{1}{2}\bm{z}' \hat G^{-1}\bm{z}'
 =v -\frac{1}{2}\bm{z} \hat G^{-1}\bm{z}
}

Because each eigenvalue $\hat G_p$ of $\hat G$ tends to $-\infty$ as
$u\tend 0$ and to $\infty$ as $u\tend\infty$, it has to vanish at
some value of $u$. On that null hypersurface, $\hat G$ becomes
singular, and the translations $\hat G \tilde{\bm{b}}_p$ in the
$z$-space are not linearly independent. If $\bm{a}=0$, the
compactified spacetime has a pathological singularity on this null
hypersurface. For example, suppose that $\hat G_1=0$ at $u=u_1$ and
$\tilde b_p^q=\delta^q_p$. Then, each point on the subspace $z^1=0$
is a fixed point of the transformation $g$ with
$\bm{a}=\bm{b}=0,\tilde{\bm{b}}=\tilde{\bm{b}}_1$. By contrast, in
the generic case in which none
of $\tilde{\bm{b}}_p$ is annihilated by $\hat G$, such a fixed point
does not exist. However, we can show that some orbits of  $\Gamma$
have accumulation points. In such cases, the quotient spacetime
obtained through compactification does not have the Hausdorff
property.

These pathological singularities can be avoided if the conditions
that $\bm{a}_p$ are generic non-vanishing vectors and different
eigenvalues of $\hat G$ vanish at different values of $u$ are both
satisfied. In this case, the vectors $(\bm{a}_p,\tilde{\bm{b}}_p)$
($p=1,\cdots,9-m$) span a $(9-m)$-dimensional subspace of the
nine-dimensional $(\bm{y},\bm{z})$-space for any value of $u>0$, and
the quotient of each $u=\const$ hypersurface is homeomorphic to
$\RF^{m+1}\times T^{9-m}$. When one of the eigenvalues of $\hat G$
vanishes, some single direction in the $\bm{z}$-space is
decompactified, but simultaneously some direction in the
$\bm{y}$-space is compactified. This implies that the direction of
the compactified extra dimensions rotates significantly with $u$ at
the point where the matrix $\hat G$ becomes singular.

\begin{figure}
\begin{center}
\includegraphics[width=12cm]{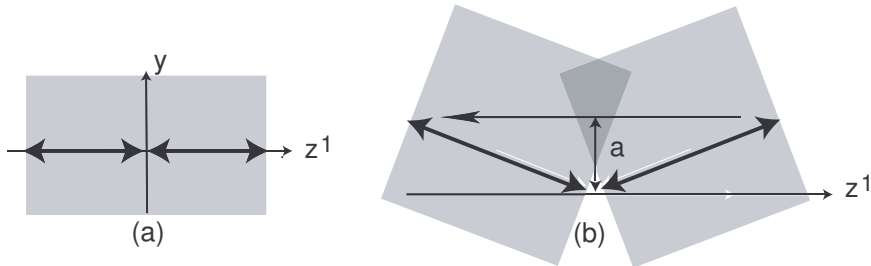}
\caption{Examples of compactification for (a) $a=0$ and (b) $a\neq 0$.
The big arrows indicate the identification, and the shaded regions represent
the corresponding fundamental domains.}
\label{fig:compact}
\end{center}
\end{figure}

The situation described above is illustrated in
Fig.~\ref{fig:compact}. We should first note that the rank of $\hat
G$ is reduced by at most 1. This implies that if we
denote the eigenvalues of $\hat G$ by $D_1(u),\cdots,D_{9-m}(u)$ and
the zero point of each $D_p(u)$ by $u_p$, then $u_p \not= u_q$ for
$p \not= q$. Suppose that $D_1 \sim k (u-u_1)$ near $u=u_1$. In the
vicinity of $u=u_1$, the dimensions $z^2,...,z^{9-m}$ are toroidally
compactified by $\hat G\tilde{\bm{b}}_2,\cdots,\hat G
\tilde{\bm{b}}_{9-m}$, which are mutually independent and
nonvanishing, while $y$ and $z^1$ are identified by the translation
\begin{eqnarray}
V:  \delta y = a,~~
\delta z^1= k(u - u_1).
\end{eqnarray}
If $a= 0$, this identification would be only for the direction
of $z^1$, and the compactification radius would vanish across
$u=u_1$, as shown in Fig.~\ref{fig:compact}(a). This gives a
singular space. However, if $a\neq0$, the compactification radius
does not vanish even for $u=u_1$, but the compactification direction
rotates in the $z^1$-$y$ space, as shown in
Fig.~\ref{fig:compact}(b). At $u=u_1$, the space decompactifies
along the $z^1$ direction but compactifies along the $y$ direction,
and in this way we can avoid the singularity.
Hence, the tilt in the compactification direction leads to a kind of
singularity resolution, and $a$ plays the role of a regularisation
parameter.
It is interesting that the compactification direction is not always
the $z^1$ direction; rather, it rotates in the direction in which
the rank of $\hat G$ is reduced. Clearly, this rotation occurs
$(9-m)$ times.

Thus, starting from a generic light-like solution with two distinct
asymptotically flat regions, as in \eqref{DLLS:AC}, we have
constructed a compactified solution that smoothly connects a static
region with constant extra dimensions to a dynamical decompactifying
region by a light-like 4-form flux. The decompactifying structure in
the limit $u\tend\infty$ can be more clearly seen in terms of the
coordinates defined by
\Eq{
\bar{\bar v}=v -\frac{1}{2}\bm{z} \hat D^{-1}\bm{z},\quad
\bar{\bar{\bm{z}}}=\hat D^{-1}\bm{z},
}
where $\hat D = u+e-e'$. In this coordinate system, the spacetime
metric for $u\sim\infty$ becomes
\Eq{
ds^2 \approx -2du d\bar{\bar v} + d\bm{y}^2
  + d\bar{\bar{\bm{z}}}\cdot \hat D^2 d\bar{\bar{\bm{z}}},
}
and the transformation $g\in\Gamma$ is written
\Eq{
   u'=u,\quad
   \bar{\bar v}'= \bar{\bar v},\quad
   \bm{y}'=\bm{y} + \bm{a},\quad
   \bar{\bar{\bm{z}}}' = \bar{\bar{\bm{z}}} + \tilde{\bm{b}}.
}
This represents a standard toroidal compactification in the
$(\bm{y},\bar{\bar{\bm{z}}})$-space.

Although we have considered only diagonal solutions here, it is
clear from the arguments given to this point that the above
construction can be easily extended to light-like solutions with  a
non-diagonal metric.
We also point out that the toroidal compactification considered in
the present section preserves 16 supersymmetries of the light-like
solution taken as the starting point. This can be explicitly
confirmed as follows.
First, under the transformation $\Phi$ given in
\eqref{DiscreteIsometry}, the frame basis $\theta^a$ transforms as
\Eqrsub{
\Phi^*\theta^- &=& \theta^-,\\
\Phi^*\theta^+ &=& \theta^+ -\frac{1}{2}\bm{h}^2 \theta^-
                   +h_I \theta^I,\\
\Phi^*\theta^I &=& \theta^I - h_I \theta^-,
}
where $h_I$ is expressed in terms of $(b_i)=(\bm{b},\tilde{\bm{b}})$ as
\Eq{
h_I= e_I^i(u) b_i.
}
{}From the general relation
\Eq{
 \Phi^* \theta^a = \Lambda^a{}_b \theta^b
 \equivalent
 S \Gamma^a S^{-1} =\Lambda^a{}_b \Gamma^b,
}
the transformation $S$ of the Killing spinor $\epsilon$ is given by
\Eq{
S=\exp\inpare{\tfrac{1}{2}h_I \Gamma^I\Gamma^-}
 =1 + \frac{1}{2}h_I\Gamma^I \Gamma^-.
}
It follows that any spinor field satisfying $\Gamma^-\epsilon=0$ is
invariant under any transformation of $G_m$. Hence, the 16
supersymmetries of the generalised light-like solution studied in \S
\ref{subsec:supersymmetry} are preserved under the toroidal
compactification considered in the present section.

\section{Summary and discussion}

In the present paper, we presented a quite general class of
spatial-coordinate independent light-like solutions with 16
supersymmetries and studied their toroidal compactification. In
particular, starting from a light-like solution which represents the
exact Minkowski spacetime in the region $u<0$ and approaches a
different Minkowski spacetime as $u\tend\infty$, we constructed a
toroidally compactified regular solution whose internal space size
is constant for $u<0$ and increases without bound as $u\tend \infty$.

The existence of such a compactification is not a trivial result,
because the Einstein equations require that if the 4-form flux $F_4$
does not vanish, the spatial metric becomes singular at a finite
value of $u$, and as a result, the standard simple toroidal
compactification produces singularities. In the present paper, we
have overcome this difficulty by considering a tilted toroidal
compactification with the help of discrete transformations with
boost components. In this compactification, the direction of the
compactified extra dimensions rotates several times when we traverse
the region with $F_4\not=0$ from one flat region to the other flat
region.

In the examples constructed in \S \ref{sec:Compactification}, the
region with static internal space occupies a half infinite region of
spacetime bounded by a null hypersurface. Hence, if we put this
region into the region $u<0$, the solution can be regarded as a
decompactification transition. However, if we apply the
time-reversal operation to this solution, we obtain a solution
describing a compactification transition. In either case, two
regions always coexist on a fixed time slice. Hence, it is difficult
to directly relate the solution to cosmology, even if we ignore the
static nature of the base spacetime.

\begin{figure}
\begin{center}
\includegraphics[width=7cm]{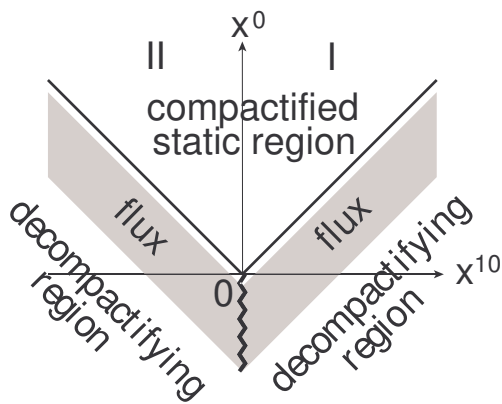}
\caption{A solution representing the creation and expansion of a
static compactified region in an asymptotically decompactifying
background.}
\label{fig:creation}
\end{center}
\end{figure}

However, we can construct an interesting solution that may be
relevant to cosmology with the following trick. We start from a
compactified solution with a static region in the region
satisfying $u=(x^0-x^{10})/\sqrt{2}>0$. We consider the region
satisfying $x^0\ge0$ and $x^{10}\ge0$ of this solution and refer to
it as I. Next, we consider the image of I under the spatial
reflection $x^{10} \maps  -x^{10}$ and refer to the corresponding
new solution in $x^0\ge0$ and $x^{10}\le0$ as II. Then, by
gluing I and II along $x^{10}=0$, we obtain a smooth compactified
solution in the region $x^0\ge0$, because the region $u>0$ of I is a
standard toroidal compactification of an exactly flat spacetime (see
Fig.~\ref{fig:creation}). In this new solution, the region with a
static internal space is given by $x^0>|x^{10}|$.
Hence, this solution represents the creation of a region with static
internal space and its subsequent expansion at the speed of light.
When this solution is extended to the region $x^0<0$, a singularity
appears. However, when quantum effects are taken into account, such
a singularity might be avoided, and in that case we would obtain a
solution describing the quantum creation of a stable compactified
region. It would be quite interesting if a similar solution with
spherical symmetry could be found in a class with a time-like
Killing spinor.

Finally, we note that the solutions presented in this paper are
expected to have a dual description in the matrix theory, which allows
a non-perturbative description of cosmological
singularities\cite{Craps.B&Sethi&Verlinde2005,Ishino.T&Ohta2006A}.
Indeed, a preliminary study of the solutions indicates that this is possible.
It would be interesting to further study the dual description of our
solutions and try to understand how the singularity may be tamed and
what kind of role the classical singularity resolution by the tilt
in the toroidal compactification plays in the dual matrix theory.

\section*{Acknowledgements}

We would like to thank C.-M. Chen for valuable discussions in the
early stages of this work.
HK  and NO were supported in part by Grants-in-Aid for Scientific
Research from JSPS (Nos. 18540265 and 16540250, respectively).

\appendix
\section{Exceptional Killing Fields for Diagonal Solutions}
\label{Appendix:ExceptionalKilling}

In this appendix, we show that the general solution to the
Killing equation in the diagonal generalised light-like
spacetime is given by \eqref{GLLS:DSCI:Killing}. We adopt
the notation used in \S \ref{subsec:ExceptionalGeometry}.

Since we are assuming that the case \eqref{GLLS:DSCI:metrictype2} does not
occur, we have
\Eq{
\kappa=\const,\quad
\eta^u=d_0+ \sum_{i\in S_1} d_i x^i.
}
Hence, for $i,j\in S_1$, \eqref{GLLS:KillingEq:reduced2} can be written
\Eq{
\pd_{i} \eta^{j}+\pd_{j}\eta^{i}
 +2(u-u_0) \pd_{i}\pd_{j} \eta^v=0.
}
This equation is equivalent to the two equations
\Eqrsub{
&& \pd_{i} \eta^{j}+\pd_{j}\eta^{i}=0,\\
&& \eta^v=\sum_{i\in S_1} \eta^v_{i}(z) x^i + \eta^v_{0}(z).}
Here and in the following $z$ represents $(x^i)_{i\in S_2}$.

Next, for $i\in S_1,\ k\in S_2$, \eqref{GLLS:KillingEq:reduced2} reads
\Eq{
e^{-2f_k}\pd_k \eta^i + \pd_i \eta^k
 +\inpare{(u-u_0)e^{-2f_k} + F_k} \pd_i \pd_k \eta^v=0.
}
Differentiation of this equation with respect to $u$ yields
\Eq{
\dot f_k \pd_k \eta^i - \inrbra{1-(u-u_0) \dot f_k}\pd_i\pd_k \eta^v=0.
}
Hence, from $K_k\not=0$, we obtain
\Eq{
\pd_k \eta^i=0,\quad
\pd_k\pd_i \eta^v=0,\quad
\pd_i \eta^k=0.
}
To summarise, for $i,j\in S_1$ we have
\Eq{
\eta^i= \zeta^i(x);\quad
\pd_{i} \zeta^{j}+\pd_{j}\zeta^{i}=0,
}
and for $k\in S_2$ we have
\Eqr{
\eta^k=\zeta ^k(z),\quad
\eta^v= \sum_{i\in S_1} B_i x^i +  \eta^v_{0}(z).
}

Finally, for $k_1,k_2 \in S_2$, \eqref{GLLS:KillingEq:reduced2} reads
\Eqr{
&& e^{2f_{k1}} \pd_{k_2}\eta^{k_1}
   + e^{2f_{k2}} \pd_{k_1}\eta^{k_2}
   +(e^{2f_{k_1}}F^{k_1}+e^{2f_{k_2}}F^{k_2})
    \pd_{k_1}\pd_{k_2} \eta^v_0
    \notag\\
&&\qquad
  -2\dot f_{k_1} e^{2f_{k_1}} \delta_{k_1\,k_2}\inpare{ \kappa u -d_0
   -\sum_{i\in S_1}d_i x^i}=0.
}
If we introduce $\tilde \zeta^k$ ($k\in S_2$) as
\Eq{
\tilde \zeta^k=\eta^k + F^k \pd_k \eta^v_0 -(\kappa u-d_0)\dot f_k x^k,
}
this equation is equivalent to $d_i=0$ [because no other term
depends on the coordinates $x^i (i\in S_1$)] and
\Eq{
\pd_{k_2} (e^{2f_{k_1}} \tilde \zeta^{k_1})
 + \pd_{k_1} (e^{2f_{k_2}} \tilde \zeta^{k_2})=0,
}
which is the Killing equation for the Euclidean space. Hence, we obtain
\Eq{
\eta^k+F^k \pd_k \eta^v_0 - (u\kappa -d_0) \dot f_k x^k
 = \sum_{l\in S_2}e^{f_l-f_k}\omega_{kl}(u) x^l + \tilde a^k(u),
}
where $\omega_{kl}=-\omega_{lk}$. Differentiating both sides of this
equation with respect to  $u$, we have
\Eq{
\pd_k \eta^v_0= 2\ell_k x^k + 2 \sum_{l\in S_2} S_{kl}x^l + a_k,
}
where
\Eqrsub{
&& 2\ell_k := e^{2f_k} \Tdot{\insbra{(\kappa u-d_0) \dot f_k}},
\label{ellk:def}\\
&& 2S_{kl}:=e^{f_k+f_l}\inrbra{\dot \omega_{kl}
  -(\dot f_k-\dot f_l)\omega_{kl}},\\
&& a_k=e^{2f_k} \Tdot{(\tilde a^k)}.
}
Since $\eta^v_0$ depends only on $z=(x^k)_{k\in S_2}$, $\ell_k,S_{kl}$ and
$a_k$ must be constant. Further, the integrability of $\pd_k \eta^v_0$
requires $S_{kl}=S_{lk}$. Therefore, we obtain
\Eqr{
&& \dot \omega_{kl}=0,\\
&& S_{kl}=-s_{kl}\omega_{kl};\quad
   2s_{kl}:= e^{f_k+f_l}(\dot f_k-\dot f_l).
}

To summarise, we have
\Eqrsub{
&& \eta^k= \sum_{l\in S_2}
    \inpare{e^{-f_k+f_l}+2s_{kl}F^k } \omega_{kl}x^l
             + e_k x^k + A_k,\\
&& \eta^v_0= C + \sum_{k\in S_2} \inpare{ B_k x^k + \ell_k (x^k)^2}
             -\sum_{k,l\in S_2} s_{kl}\omega_{kl} x^k x^l,
}
where
\Eq{
e_k=(\kappa u_0 - d_0) \dot f_k(u_0).
}
Here, since both $S_{kl}$ and $\omega_{kl}$ are constant,  the
relation $\omega_{kl}\not=0$ holds only when $s_{kl}=-s_{lk}$ is
constant.
Inserting these expressions into \eqref{GLLS:SCI:Killing}, we obtain
the result, \eqref{GLLS:DSCI:Killing}.


\end{document}